\documentclass[proceedings]{JHEP3}

\PrHEP{PrHEP hep2001}                   
\conference{International Europhysics Conference on HEP}                

\usepackage{epsfig}			

\newcommand\sss{\scriptscriptstyle}

\def\as{\alpha_{\sss S}}

\def\ord{{\cal O}}

\def\({\left(} 
\def\){\right)} 

\def\beq{\begin{equation}}
\def\eeq{\end{equation}}
\def\beqa{\begin{eqnarray}}
\def\eeqa{\end{eqnarray}}
\def\lsim{\mathrel{\raisebox{-.6ex}{$\stackrel{\textstyle<}{\sim}$}}}

\newbox\mybox
\newcommand\fverb{\setbox\mybox=\hbox\bgroup\verb}
\newcommand\fverbdo{\egroup\medskip\noindent\fbox{\unhbox\mybox}\ }
\newcommand\fverbit{\egroup\item[\fbox{\unhbox\mybox}]}

\font\beeg=cmr17 scaled 1600		
\newcommand\init[1]{\setbox\mybox=\hbox{{\beeg #1}~}%
		   \noindent\global\hangindent=\wd\mybox\global\hangafter-2%
		   \sc\smash{\llap {\lower 13.2pt \box\mybox}}}

\title{Higgs production plus two jets at hadron colliders}

\author{\speaker{V. Del Duca}\\
	I.N.F.N., Sez. di Torino, via P. Giuria, 1 - 10125 Torino, Italy\\
        E-mail: \email{delduca@to.infn.it}}

\author{W. Kilgore\\
Physics Department, Brookhaven National Laboratory\\
Upton, New York 11973, U.S.A.\\
	E-mail: \email{kilgore@bnl.gov}}

\author{C. Oleari and D. Zeppenfeld\\
Department of Physics, University of Wisconsin\\
 Madison, WI 53706, U.S.A.\\
	E-mail: \email{oleari@pheno.physics.wisc.edu, 
dieter@pheno.physics.wisc.edu}}

\author{C.R. Schmidt\\
Department of Physics and Astronomy,
Michigan State University\\
East Lansing, MI 48824, USA\\
	E-mail: \email{schmidt@pa.msu.edu}}

\abstract{In this talk we present a calculation of Higgs production via
gluon fusion in association with two jets, including the full top-quark mass
dependence, and compare it to the large top-mass limit. We find that
the large top-mass limit is a good approximation as long as the Higgs
mass is smaller than the top quark pair mass, and the jet transverse energies
are smaller than the top mass. In addition, we compare Higgs production 
via gluon fusion and via weak-boson fusion,
and consider final-state distributions, like the rapidity
interval between the jets and the jet-jet azimuthal decorrelation,
which may allow us to distinguish one fusion process from the other.
}

\begin{document} 

\section{Higgs production at hadron colliders}

At the Large Hadron Collider (LHC) a Higgs boson is expected to be 
produced mainly by gluon fusion 
or weak-boson fusion (WBF)~\cite{CMS,ATLAS}. 
The WBF channel, even though numerically 
smaller, is interesting because it is expected to provide information
on Higgs boson couplings~\cite{Zeppenfeld:2000td}. In addition, it is 
theoretically simple to analyse, since it produces the Higgs boson 
via $t$-channel $W$ or $Z$
exchange, in association with two forward quark jets, $qQ\to qQH$.
QCD radiative corrections to WBF are known to be 
small~\cite{WBF_NLO}~\footnote{The $t$-channel singlet exchange ensures 
that at next-to-leading order (NLO) no gluon
is exchanged in the $t$-channel, unless the incoming quarks are of
equal flavour.}, and thus this process promises small systematic errors.

The gluon-fusion channel is much more challenging.
The Higgs boson couples to gluons via a quark loop, thus every calculation 
to a given loop accuracy in QCD implies one more heavy-quark loop.
Since the scattering amplitude is proportional to the quark mass 
squared~\footnote{The Yukava coupling is proportional to $m_q$, and there 
is an additional factor of $m_q$ due to the compensation of the chirality 
flip, induced by the insertion of a single scalar $Hq\bar q$ vertex.}, in the
numerical evaluation of the production rate it suffices to consider only
the top quark. 
Inclusive Higgs production via gluon fusion, $g\, g
\to H$, is known at NLO in QCD, including the
full $m_t$ dependence~\cite{HggNLO}, and the NLO corrections are known 
to be large ($\ord(100\%)$). At next-to-next-to-leading order (NNLO), 
inclusive Higgs production via 
gluon fusion is known only in the $m_t\to\infty$
limit~\cite{H2loop}, for which the top-quark loop reduces to an 
effective $H\,g\,g$ coupling, and the calculation reduces from three to 
two loops. However, in the 
intermediate Higgs mass range, which is favoured by electroweak precision
data~\cite{LEPEWWG}, the Higgs boson mass $m_H$ is small compared to the 
top-quark pair threshold and the large $m_t$ limit promises to be an 
adequate approximation.

$H + 1$ jet production via gluon fusion is known at leading order (LO), 
including the
full $m_t$ dependence~\cite{Ellis:1988xu}, and at NLO in the $m_t\to\infty$
limit~\cite{deFlorian:1999zd}. Also in this case the NLO corrections are known 
to be large, roughly of the same order as in the inclusive case.
$H + 2$ jet production via gluon fusion was previously known at LO, 
but only in the $m_t\to\infty$ limit~\cite{Dawson:1992au}.
Both in $H + 1$ jet and in $H + 2$ jet production, phase space regions 
open up where one or several of the kinematical invariants  
are of the order of, or exceed, the top-quark mass, i.e.\ regions of 
large Higgs boson or jet transverse momenta, or regions where dijet invariant  
masses become large. These regions may invalidate the $m_t\to\infty$
limit, even in the intermediate Higgs mass range, but 
yield a rather small contribution to inclusive Higgs production.
However, in $H + 1$ jet or in $H + 2$ jet production one may require
a Higgs-jet mass or a dijet mass to be large (e.g. in $H + 2$ jet 
production we may require the dijet mass to be large, 
in order to compare with the
large dijet mass of the two forward jets in the WBF process), then
it is important to evaluate the extent to which the $m_t\to\infty$
limit holds. Therefore, in this talk we review the results of a 
recent calculation of
$H + 2$ jet production, including the full $m_t$ dependence~\cite{dkosz},
where the issues above are analysed.

\DOUBLEFIGURE[t]{ggh_no_cuts.eps, width=.4\textwidth}
{ggh_cuts.eps, width=.4\textwidth}{\label{fig:sigmaMh} 
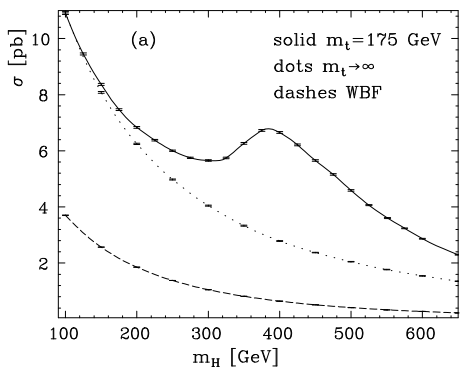
$H+2$~jet cross sections in pp collisions at
{\rm $\sqrt{s}=14$~TeV} as a function of the Higgs boson mass.
Results are shown for gluon-fusion processes induced by a top-quark loop
with {\rm $m_t=175$~GeV} and in the $m_t\,\to\,\infty$ limit, 
and for weak-boson fusion. The cuts of Eq.~(2.1)
have been used.}{\label{fig:sigmaMhb} 
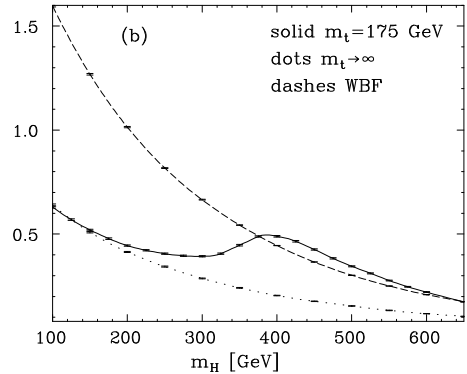
Same as Fig.~1, but with the WBF selection of 
cuts of Eqs.~(2.1) and (2.2)}

\section{Higgs production plus two jets at the LHC}

Gluon fusion and weak-boson fusion 
($qQ\,\to\, qQH$ production via $t$-channel exchange of a $W$ or $Z$), 
are expected to be the dominant sources of $H+2$~jet events at the LHC. The
impact of the former on LHC Higgs phenomenology is determined by the relative 
size of these two contributions. We evaluate the $H+2$~jet cross section
through a  minimal set of cuts on the final-state partons, which
anticipates LHC detector capabilities and jet finding algorithms,
\beq
\label{eq:cuts_min}
p_{j\perp}>20\;{\rm GeV}, \quad |\eta_j|<5,\quad R_{jj}>0.6\, ,
\eeq
where $p_{j\perp}$ is the transverse momentum of a final state parton
and $R_{jj}$ describes the separation of the two partons in the 
pseudo-rapidity $\eta$ versus azimuthal angle plane
$R_{jj} = \sqrt{\Delta\eta_{jj}^2 + \phi_{jj}^2}$.
Expected $H+2$~jet cross sections at the LHC are shown in
Fig.~\ref{fig:sigmaMh}, as a function of the Higgs boson mass, $m_H$.  The
three curves compare results for the expected Standard Model gluon-fusion 
cross section
at $m_t=175$~GeV (solid line) with the large-$m_t$ limit (dotted line), 
computed using the heavy-top effective Lagrangian, and
with the WBF cross section (dashed line).
Error bars indicate the statistical errors of the Monte Carlo integration. 
Cross sections correspond to the sum over all Higgs decay modes: finite Higgs
width effects are included. The factorization scale
was set to $\mu_f=\sqrt{p_{1\perp} \, p_{2\perp} }$, and
$\as$ was taken to be $\as(M_Z) =0.12$. Different choices of
renormalization and factorization scales have been discussed in 
Ref.~\cite{dkosz}, where a strong sensitivity of the $H+2$~jet 
cross section on the renormalization scale was found.

Fig.~\ref{fig:sigmaMh} shows cross sections within the
minimal cuts of Eq.~(\ref{eq:cuts_min}). The gluon-fusion contribution 
dominates because the cuts retain events with jets in the central region, 
with relatively small dijet invariant mass. In order to assess background 
levels for WBF events, it is more appropriate to consider typical tagging 
jet selections employed for WBF studies~\cite{RZ_WW}. This is done in 
Fig.~\ref{fig:sigmaMhb} where, in addition to the cuts of 
Eq.~(\ref{eq:cuts_min}), we require
\beq 
\label{eq:cut_gap}
|\eta_{j1}-\eta_{j2}|>4.2\, , \qquad \eta_{j1}\cdot\eta_{j2}<0\, ,
\qquad m_{jj}>600\;{\rm GeV}\, ,
\eeq
i.e.\ the two tagging jets must be well separated, they must 
possess a large dijet invariant mass, and must
reside in opposite detector hemispheres.
With these selection cuts the weak-boson 
fusion processes dominate over gluon fusion by about 3/1 for 
Higgs boson masses in the 100 to 200~GeV range. This means that a relatively
clean separation of weak-boson fusion and gluon-fusion processes will be 
possible at the LHC, in particular when extra central-jet-veto techniques are 
employed to further suppress semi-soft gluon radiation in QCD 
backgrounds~\cite{RZ_WW}.

\EPSFIGURE[ht]{phi_cuts_comp.eps,width=7cm} {
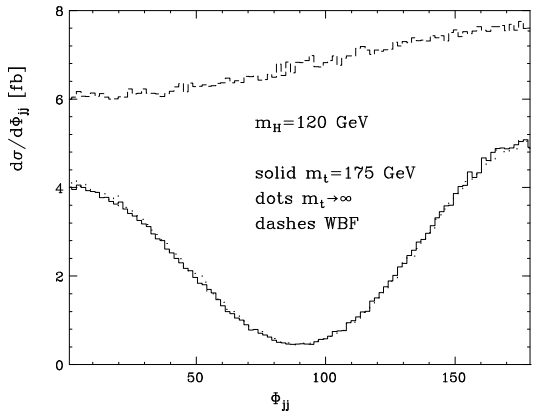
\label{fig:phi_cuts_comp} Azimuthal-angle distribution between
the two final jets, with the WBF cuts of Eqs.~(\ref{eq:cuts_min})
and (\ref{eq:cut_gap}). 
Results are shown for gluon-fusion processes induced by a top-quark loop
with $m_t=175$~{\rm GeV} and in the $m_t\,\to\,\infty$ limit, computed 
using the heavy-top effective Lagrangian, and for weak-boson fusion. }

A conspicuous feature of the $H+2$~jet gluon-fusion cross sections in 
Figs.~\ref{fig:sigmaMh} and \ref{fig:sigmaMhb} is the threshold 
enhancement at $m_H\approx 2\,m_t$,
an effect which is familiar from the inclusive gluon-fusion cross section.
Well below the threshold-peak
region, the large $m_t$ limit provides an excellent 
approximation to the total $H+2$~jet rate from gluon fusion, at least when 
considering the total Higgs production rate only. 
Fig.~\ref{fig:sigmaMhb} also implies that the
approximation provided by the large $m_t$ limit at Higgs boson masses below
about 200~GeV is excellent. Thus the large dijet invariant mass,
$m_{jj}>600$~GeV, and the concomitant large parton center-of-mass energy 
do not spoil the $m_t\,\to\,\infty$ approximation. A hint to understand that
can be found in the high-energy limit, which is appropriate for the large
dijet mass case. In the high-energy limit
only Feynman diagrams with gluon exchange in the $t$ channel 
are relevant. When taking in addition the large $m_t$ limit, the same
diagrams contribute. Thus, if the high-energy limit is appropriate to
describe the region of large dijet mass, so is the combined high-energy 
and large $m_t$ limit. The only difference is in the high-energy 
coefficient functions, or impact factors, which in the combined limit
lose the information on the $m_t$ dependence.
Finally, as shown in Ref.~\cite{dkosz}, the large $m_t$ limit works well 
in the intermediate Higgs mass range, as long as jet 
transverse momenta stay small: $p_{j\perp}\lsim m_t$.

Turning now to the issue of differentiating between
gluon fusion and WBF processes, 
a characteristic of WBF events is the large rapidity separation 
of the two tagging jets, a feature which is not shared by $H+2$~jet events 
arising from gluon fusion. The plots of the rapidity separation of the 
jets~\cite{dkosz} show indeed that for the inclusive cuts of 
Eq.~(\ref{eq:cuts_min}) jets coming from WBF events are produced
preferentially with a rather large rapidity separation, while jets
coming from gluon fusion events are produced mostly in the central 
rapidity region. Accordingly, when WBF cuts (\ref{eq:cuts_min}) and
(\ref{eq:cut_gap}) are implemented the jets coming from 
gluon fusion events are depleted, thus the jet separation cut
is one of the most effective means of enhancing WBF processes with respect to
gluon fusion.

Another jet-angular correlation, which allows to distinguish
gluon fusion from weak-boson fusion, is the azimuthal angle between the two
jets, $\phi_{jj}$. The distributions for gluon-fusion and WBF processes are
shown in Fig.~\ref{fig:phi_cuts_comp}. 
In the WBF process $qQ \,\to\, qQH$, the matrix element squared 
$|{\cal A}_{\rm WBF}|^2$ is proportional to ${\hat s} m_{jj}^2$, with
${\hat s}$ the squared parton center-of-mass energy and $m_{jj}$ the
dijet invariant mass.
Since the dependence of $m_{jj}^2$ on $\phi_{jj}$ is mild, we have the flat
behavior depicted in Fig.~\ref{fig:phi_cuts_comp}.
The azimuthal-angle distribution of the gluon-fusion process is instead
characteristic of the CP-even operator $H G_{\mu\nu}G^{\mu\nu}$, where
$G_{\mu\nu}$ is the gluon field strength tensor. 
This effective coupling
can be taken as a good approximation for the $ggH$ coupling in the 
large-$m_t$ limit. Infact the large-$m_t$ limit (dotted line) is almost
indistinguishable from the $m_t=175$~GeV result (solid line).
Not only does the $\phi_{jj}$ correlation allow us to distinguish
WBF from gluon fusion, it can also be used as a tool to investigate
the tensor structure of the $WWH$ coupling. Infact, if we suppose that
there is an anomalous (i.e.\ non Standard Model) $WWH$ coupling, which in
the low-energy effective theory can be modeled through higher-dimensional
operators, the $\phi_{jj}$ correlation discriminates between a 
CP-even coupling, which behaves like the $ggH$ coupling, and a
CP-odd coupling, which would hinder configurations where the jets
are aligned or back-to-back~\cite{PRZ}.


\begin{thebibliography}{99}

\bibitem{CMS}
G.~L.~Bayatian {\it et al.}, CMS Technical Proposal,
report CERN/LHCC/94-38x (1994);
R.~Kinnunen and D.~Denegri, 
CMS NOTE 1997/057;
R. Kinnunen and A. Nikitenko,
CMS TN/97-106;
R.~Kinnunen and D.~Denegri,
[\hepph{9907291}].

\bibitem{ATLAS}
ATLAS Collaboration, ATLAS TDR,
report CERN/LHCC/99-15 (1999).

\bibitem{Zeppenfeld:2000td}
D.~Zeppenfeld, R.~Kinnunen, A.~Nikitenko and E.~Richter-Was,
\prd{62}{2000}{013009} [\hepph{0002036}].

\bibitem{WBF_NLO}
T.~Han and S.~Willenbrock, \plb{273}{1991}{167}.

\bibitem{HggNLO}
A.~Djouadi, N.~Spira and P.~Zerwas, \plb{264}{1991}{440};
M.~Spira, A.~Djouadi, D.~Graudenz and P.M.~Zerwas,
\npb{453}{1995}{17} [\hepph{9504378}];
S. Dawson, \npb{359}{1991}{283}.

\bibitem{H2loop}
S.~Catani, D.~de Florian and M.~Grazzini, \jhep{05}{2001}{025}
[\hepph{0102227}];\\
R.~Harlander and W.~Kilgore, \prd{64}{2001}{013015} [\hepph{0102241}]. 

\bibitem{LEPEWWG}
See e.g.\ LEP Electroweak Working Group, LEPEWWG Note 2001-01 (2001) and
http://lepewwg.web.cern.ch/LEPEWWG/.

\bibitem{Ellis:1988xu}
R.~K.~Ellis, I.~Hinchliffe, M.~Soldate and J.~J.~van der Bij,
\npb{297}{1988}{221}.

\bibitem{deFlorian:1999zd}
D.~de Florian, M.~Grazzini and Z.~Kunszt,
\prl{82}{1999}{5209}
[\hepph{9902483}].

\bibitem{Dawson:1992au}
S.~Dawson and R.~P.~Kauffman,
\prl{68}{1992}{2273};\\
R.~P.~Kauffman, S.~V.~Desai and D.~Risal,
\prd{55}{1997}{4005} 
[\hepph{9610541}].

\bibitem{dkosz}
V.~Del Duca, W.~Kilgore, C.~Oleari, C.~Schmidt and D.~Zeppenfeld,
[{\tt hep-ph/0105129}];
[{\tt hep-ph/0108030}].

\bibitem{RZ_WW}
D.~Rainwater, R.~Szalapski and D.~Zeppenfeld,
\prd{54}{1996}{6680};
D.~Rainwater and D.~Zeppenfeld,
\prd{60}{1999}{113004}, Erratum-ibid. \prd{61}{2000}{099901};
D.~Rainwater, PhD thesis, [\hepph{9908378}];
T.~Plehn, D.~Rainwater and D.~Zeppenfeld,
\prd{61}{2000}{093005}.

\bibitem{PRZ} 
T.~Plehn, D.~Rainwater and D.~Zeppenfeld,
[{\tt hep-ph/0105325}].

\end{thebibliography}
\end{document}